\documentclass{aa}
\usepackage{graphicx}
\usepackage{txfonts}
\usepackage{natbib}
\usepackage{longtable}

\begin{document}
\def\logg{$\log (g)$~}
\def\Teff{$T_{\rm eff}$~}
\def\FeH{$[Fe/H]$~}
\def\vmicc{$\xi_{t}$}
\def\vmic{$\xi_{t}$~}
\def\msun{$M_{\odot}$~}

\title{Does the chemical signature of TYC 8442$-$1036$-$1 originate from a
  rotating massive star that died in a faint explosion?\thanks{Based on observations made with the ESO Very Large Telescope
at Paranal Observatory, Chile (ID 094.B-0781(A); P.I. G. Cescutti).}}

\author {G. Cescutti\inst{1,2} \thanks {email to: g.cescutti@herts.ac.uk} 
\and  M. Valentini \inst{3} \and P. Fran\c cois \inst{4,5}\and C. Chiappini \inst{3}
\and E. Depagne\inst{6} \and N. Christlieb\inst{7} 
\and C. Cort\'es \inst{8,9} }
\institute{Centre for Astrophysics Research, School of Physics, Astronomy and
Mathematics, University of Hertfordshire, College Lane, Hatfield AL10
9AB, UK
\and BRIDGCE UK Network (www.bridgce.net), UK
\and Leibniz-Institut f\"ur   Astrophysik Potsdam (AIP), An der Sternwarte 16, 14482, Potsdam,  Germany 
\and GEPI - Observatoire de Paris, 64 avenue de l'Observatoire, 75014, Paris, France
\and Universit\'e de Picardie Jules Verne, 33 rue St-Leu, 80080 Amiens, France
\and South African Astronomical Observatory (SAAO), Observatory Road Observatory Cape Town, WC 7925, South Africa
\and Zentrum f\"ur Astronomie der Universit\"at Heidelberg, Landessternwarte, K\"onigstuhl 12, 69117 Heidelberg, Germany
\and Departamento de F\'isica, Facultad de Ciencias B\'asicas,
  Universidad Metropolitana de la Educaci\'on, Av. Jos\'e Pedro Alessandri
  774, 7760197 Nu\~noa, Santiago, Chile
\and Millennium Institute of Astrophysics (MAS), Santiago, Chile
}

\date{Received xxxx / Accepted xxxx}

\abstract {We have recently investigated the origin of chemical
  signatures observed in Galactic halo stars by means of a stochastic
  chemical evolution model.  We have found that rotating massive stars
  are a promising way to explain several signatures observed in these
  fossil stars.}{In the present paper we discuss how the extremely
  metal-poor halo star TYC 8442$-$1036$-$1, for which we have now
  obtained detailed abundances from VLT-UVES spectra, fits into the
  framework of our previous work.}{We apply a standard 1D LTE analysis
  to the spectrum of this star. We measure the abundances of 14
  chemical elements; for Na, Mg, Ca, Sc, Ti, V, Cr, Mn, Fe, Ni and Zn we
  compute the abundances using equivalent widths; for C, Sr and Ba we
  obtain the abundances by means of synthetic spectra generated by
  MOOG.}{We find an abundance of [Fe/H]=~$-$3.5~$\pm0.13$ dex based on
  our high resolution spectrum; this points to an iron content lower
  by a factor of three (0.5 dex) compared to the one obtained by a low
  resolution spectrum.  The star has a [C/Fe]~=~0.4 dex, and it is not
  carbon enhanced like most of the stars at this metallicity.
  Moreover, this star lies in the plane [Ba/Fe] vs. [Fe/H] in a
  relatively unusual position, shared by a few others galactic halo
  stars that is only marginally explained by our past results.}{The
  comparison of the model results with the chemical abundance
  characteristics of this group of stars can be improved if we
  consider in our model the presence of faint supernovae coupled with
  rotating massive stars. These results seem to imply that 
  rotating massive stars and faint supernovae scenarios are complementary to each
  other, and are both required in order to match the observed
  chemistry of the earliest phases of the chemical
  enrichment of the Universe.}


\titlerunning{A new extremely metal-poor stars}

\authorrunning{Cescutti et al.}

\maketitle

\section{Introduction}
The study of extremely metal-poor (EMP) stars is of fundamental importance to
reveal the nucleosynthesis production of the first stars and how they formed,
and more in general they can be of high value to understand the
behaviour of all massive stars. Therefore, the last 20
years have seen  incredible efforts by observers worldwide to measure
these elusive, far and faint objects in the Galactic halo, from
the pioneering studies of \citet{MCW95} and \citet{Ryan96} to a new
generation of data with 8m telescopes as e.g. in \citet{Cayrel04} and
\citet{Aoki05}, to the most recent  works e.g. by \citet{Yong13} and
\citet{Roederer14b}.

 In our recent work, we have provided an interpretation to the presence in
EMP stars of specific chemical signatures by means of stochastic
chemical evolution models.  Our results supported the scenario in
which the first stars that exploded and polluted the pristine
interstellar medium (ISM) were rotating faster than the present day
massive stars. Stellar evolution codes coupled with nuclear reaction
chains have shown that this rotation produces mixing in the interior
of the stars.  This mixing impacts the nucleosynthesis of light
elements such as carbon, nitrogen and oxygen
\citep{Hirschi07,Meynet06} and it also predicts the production of
s-process elements \citep{Pigna08,Frisch12,Frisch15}. In this scenario
in which the stars were fast rotating, chemical evolution models
were able to explain several chemical anomalies observed in the early
Universe: the almost Solar ratio of [N/O] and the increase and the
spread in the [C/O] ratio \citep{Chiappini06}, the low
$^{12}$C/$^{13}$C ratios \citep{Chiappini08}, the spread present in
the [C/O] and in the [N/O] ratios \citep{Cesc10}, the primary
evolution of Be and B \citep{Prantzos12}, the spread between light and
heavy neutron capture elements \citep{Cescutti13}.

In \citet{Cescutti14}, we also predicted that in this scenario, EMP
stars with a supersolar [Sr/Ba] ratio were expected to have a barium
manly composed by even isotopes, clear signature of an s-process
pollution by fast rotating massive stars.  The observations that we
present here, were granted in the context of the ESO proposal
``Probing the sources of synthesis of neutron capture elements:
Isotopic ratios of barium in halo stars'', meant to verify this
thesis.  However, the metal-poor halo stars TYC 8442$-$1036$-$1 was a
back-up target and it was selected on the criterium to be an
(extremely) metal-poor star in the correct position in the sky, never
observed with high-resolution spectroscopy and was observed due to bad
weather conditions that prevented us to observe our main targets.  TYC
8442$-$1036$-$1 was not a striking case among the 1777 bright
(9~$<$~B~$<$~14) metal-poor candidates selected from the Hamburg/ESO
Survey (HE 2220$-$4840) by \citet{Frebel06}. Its iron content measured
from analyses of medium-resolution follow-up spectroscopy was
[Fe/H]=~$ -$2.91 dex, quite low but not so impressing in a survey that
managed to find HE 1327$-$2326 - the most iron-poor star for almost a
decade \citep{Frebel05}, only recently replaced in this record by SMSS
J031300.36$-$670839.3 \citep{Keller14}.

Nevertheless, our detailed abundance analysis presents surprises,
among which the star is more metal-poor that what expected by the
medium resolution results. The star presents also a [C/Fe]~=~0.4 dex
and  therefore is not a carbon enhanced star, like most of the
  stars at this metallicity \citep{Placco14}; actually, it looks more
  like a normal star similar to others studied by the First Stars
  collaboration \citep{Cayrel04, Spite05, Spite06}.  TYC
  8442$-$1036$-$1 shows also chemical characteristics in the [Ba/Fe]
  vs [Fe/H] space that are just at the edge of predictions for our
  best model results in \citet{Cescutti14}.  The case of TYC
  8442$-$1036$-$1 is relative rare but not unique, in fact about other
  20 stars are just marginally consistent with our previous modelling;
  moreover, few other stars ($\sim$2$-$3) cannot be explained by our
  model. Among the uncertainties of this model, there is the
scenario of the r-process events, which is the magneto-rotationally
driven scenario in \citet{Cescutti14}.  However, a similar outcome is
obtained by assuming different scenarios for the r-process events, as in
\citet{Cescutti13} by adopting the electron-capture scenario and in
\citet{Cescutti15} by adopting the neutron stars merger scenario. 

In the recent years, another scenario has been investigated to explain
the characteristics of the EMP (and ultra metal-poor stars) of the
halo.  This new scenario does not refer to characteristics of the
stars during their lives, but rather on their explosions
\citep{Cooke14}. In fact, they investigate the impact of a variations
in the explosion energy of the primordial SNe, as suggested by
\citet{Tominaga07} in the scenario of the faint SNe.  The faint SNe
produce less Fe compared to normal SNe and impacting in the results of
a stochastic chemical evolution model for the early Universe.
We decide to investigate in this paper also this scenario to see
the impact to the neutron capture elements, not considered in the
previous studies, and to compare its results to
 the results provided by the scenario of rotating massive stars. 
Moreover, the different explosion energy does not impact the
production of the studied chemical elements determined by fast
rotation.  Therefore,  the two scenario can be complementary and 
 we will show results in which the two scenarios are
coupled; these results can represent a solution to the class of 
objects with chemical characteristics similar to TYC 8442$-$1036$-$1.

\section {Observations and data reduction}

The observations were performed using UVES high-resolution
spectrograph \citep{Dekker00} in slit mode, mounted at the UT2 Kueyen
Telescope at the ESO Paranal Observatory (Chile).  We adopted the
standard R530 setup (Red Arm only, Cross Disperser 3 and centered at
520 nm), covering the wavelength interval 4140--6210\,$\AA$, with a
resolving power of R~$\sim$100,000.

\begin{table*}[ht!]
\begin{minipage}{190mm}
\caption{Observing log}

\begin{tabular}{|c|c|c|c|c|c|c|c|c|c|}
  \hline
  Object  & R.A.                 & Decl. & B & V & Exposure & S/N & Obs. Date & $V_{r}$& error  \\
             &  (J2000.0)         & (J2000.0) & (mag) & (mag) & time (s) & &   (UT)  & ($km s^{-1})$ & ($km s^{-1})$\\
                                                                 
 \hline
TYC 8442$-$1036$-$1 &  22 23 23.3 & -48 24 50.9 & 11.71 & 11.10&  1200s & $>$80 &4th October 2014 & 90.6 & 0.6  \\
\hline
\end{tabular}
\label{tablog}
\end{minipage}
\end{table*}

  The spectrum was acquired on the night of 4th October 2014,
  adopting an exposure time of 20 minutes.  The resulting spectrum has
  an average SNR of 100 (from 80 at 4200\,$\AA$ to 150 at $\lambda>$~5500\,$\AA$
  ). Details of the observation are summarised in Table~\ref{tablog}.

  The spectrum was reduced using the ESO standard pipeline for
  UVES. The radial velocity (RV) was measured from the H-$\beta$ line.
  The RV correction was applied using the standard IRAF package
  \textit{RVcorrect}.  The correction was confirmed by the subsequent
  comparison with the wavelengths of Fe and Ca lines.

\section {Atmospheric parameters and abundances}

\begin{figure*}[ht!]
\begin{minipage}{185mm}

\includegraphics[width=185mm]{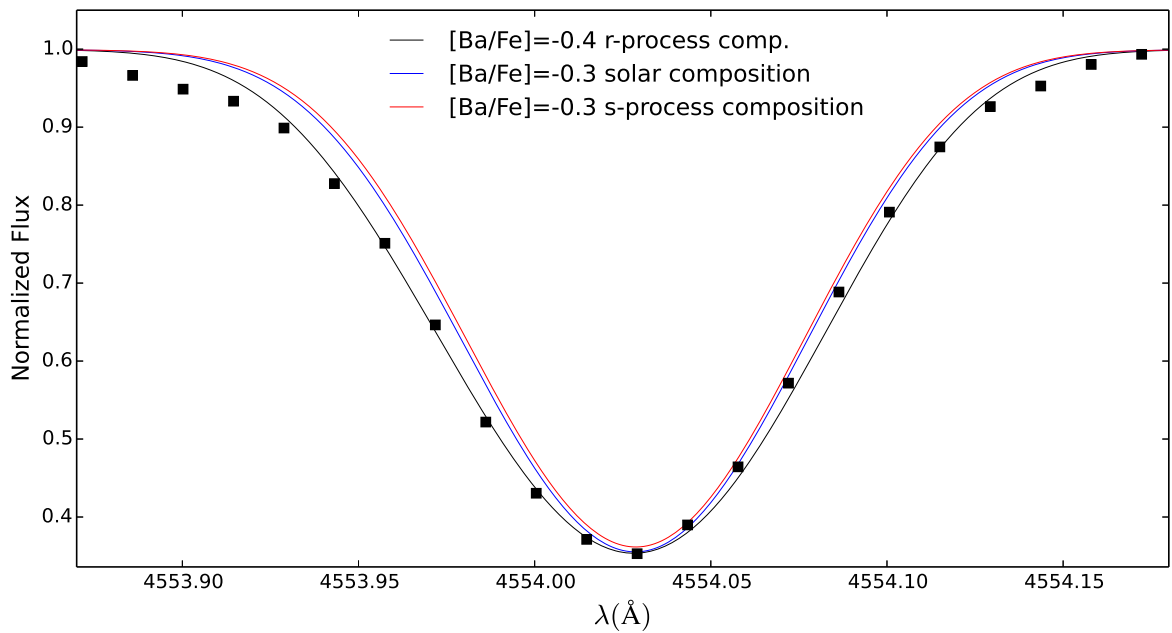}

\caption{Fit of the Ba line at 4554~$\AA$. The observed spectrum are represented by dots, the synthetic spectra by lines.
 We present the different results obtained  using different isotopic compositions for Ba:
 r-process composition, s-process composition and solar composition. 
Note that we have obtained different  best abundances for the different compositions.}\label{fig1c}
\end{minipage}
\end{figure*}

\begin{figure*}[ht!]
\begin{minipage}{185mm}
\includegraphics[width=185mm]{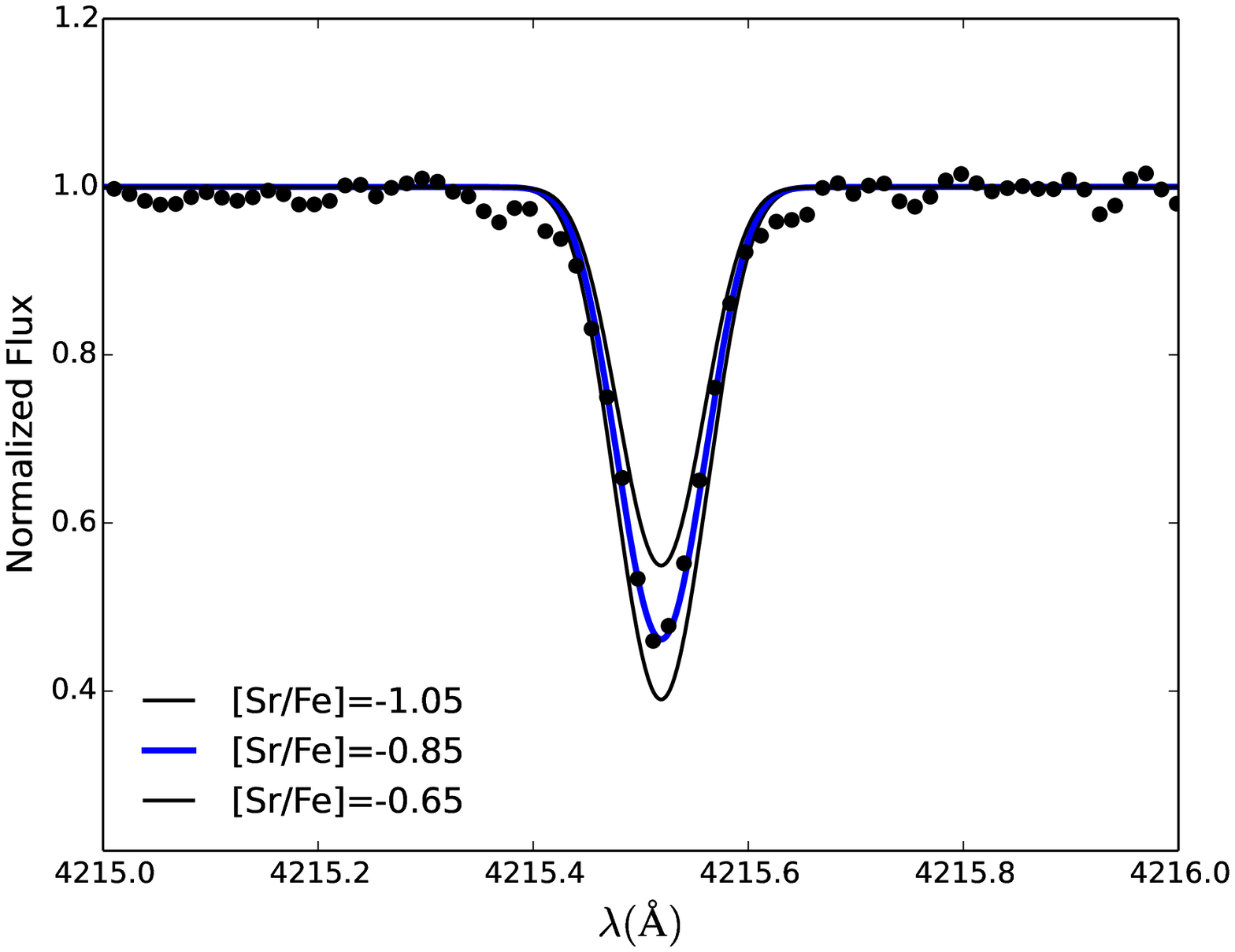}

\caption{Fit of the Sr line at 4607\,$\AA$. Dots:  observed spectrum, lines synthetic spectra. 
The synthetic spectra are calculated for 3 different abundances of Sr. The intermediate one 
has been taken as best value for the Sr abundance measured in this star.}\label{fig1b}
\end{minipage}
\end{figure*}

We use stellar model atmospheres interpolated
from the grid of one-dimensional MARCS models  \citep{Gustafsson08} 
and performed the analysis using a recent version of the spectral
line analysis code MOOG
\footnote{http://www.as.utexas.edu/$\sim$chris/moog.html}
. Effective temperatures (\Teff) and microturbulent velocities
( \vmicc) are derived by requiring that abundances derived from
Fe I lines showed no trend with the excitation potential and line
strength. The \logg  is derived by
requiring that the Fe abundance derived from Fe I lines matched that
derived from Fe II lines. This analysis technique is a standard analysis
and very similar to the one adopted in \citet{Roederer14}. 
Our calculations assume local
  thermodynamical equilibrium (LTE). 
The reference Solar abundances
  used in this work are taken from \citet{Asplund09}.

  The line list adopted in this work has been created starting from
 the line  adopted in \citet{Hill02},~\citet{Roederer13} and \citet{Ural15} 
  supplemented by lines taken directly from VALD3 \citep{Kupka00}.  The line list
  was constructed in order to avoid as much as possible blends with
  other atomic lines and molecular bands.  The EW of the lines have
  been measured using the ARES \citep[Automatic Routine for
  line Equivalent widths in stellar Spectra;][]{Sousa07}. In Table~ \ref{tab:ew1}, we report 
  the wavelengths, excitation energy of the lower energy level,
  oscillator strength and EW for the considered lines.
 A fraction of the considered lines has been measured using standard
  IRAF routine: no bias or significative differences between the
  automatically measured EWs and the ones determined with the
  automatic routine have been detected.
The final atmospheric parameters derived are \Teff~=~4800K, 
 \logg~=~1.71 dex and [Fe/H]=~$-$3.50 dex
with  \vmic=~1.71 km s$^{-1}$.

\begin{table*}[ht!]
\begin{minipage}{180mm}
\caption{Abundances of chemical elements in TYC 8442$-$1036$-$1. 
For the abundance values in column 4, we use the Solar 
abundances by \citet{Asplund09}. }

\begin{tabular}{|c|c|c|c|c|c|c|c|c|c|}
\hline
chemical elements& $\epsilon(X)$ &[X/H] & [X/Fe] & $\sigma_{\epsilon(x)}$&  N. lines & random &  systematic & total    & corrections non-LTE   \\
&                      &         &             &                                 &              &   error  &   error          &   error  & \\
\hline
C (CH) &    5.33  & $-$3.10  &   0.4       &    --     &  synth  &(0.20)&   (0.12)  & 0.23& -\\    
Na I    &    2.48   &$-$3.76   &$-$0.26  & 0.01     &    2      & 0.11 &  0.03 & 0.11 & $-$0.06\\
Mg I    &    4.73   & $-$2.88 & 0.62       & 0.05     &    3      & 0.11 & 0.01 & 0.11 & 0.10 \\
Ca I    &     3.14   &$-$3.20  & 0.30       &  0.07    &  11     & 0.07 &  0.05 & 0.08 & -\\
Sc II    &$-$0.27  &$-$3.42  & 0.08       &  0.06    &    6     & 0.06 &0.11 &  0.12&-\\
Ti I     &     1.67   &$-$3.27  & 0.23       &  0.07    &    8     & 0.07 & 0.03 & 0.08&-\\
Ti II    &     1.65   &$-$3.29  & 0.21       &  0.10    &  23     & 0.10 & 0.12 & 0.15&-\\
V I     &     0.42    &$-$3.51  & $-$0.01 &  --       &    1     & 0.11 & 0.04 & 0.12&-\\
Cr I   &     1.63    & $-$4.01  &$-$0.51  & 0.09     &   4      & 0.11 & 0.03 & 0.11& -\\
Mn I  &     1.79    & $-$3.64  &$-$0.14  & 0.07     &    2     & 0.11 & 0.03 & 0.11& - \\
Fe I   &     4.00    &$-$3.50   &  --         & 0.06     &   86    & 0.06 & 0.11  & 0.13& 0.10\\
Fe II  &     4.00    &$-$3.50   &   --        & 0.11     &   11    & 0.11 & 0.09  & 0.14& -\\
Ni I   &    3.03     & $-$3.19  & 0.31       & --        &  1      & 0.11  &  0.04  & 0.12 & -\\
Zn I   &    1.36     & $-$3.20  & 0.30       & 0.09     &  2      & 0.11  &  0.05  & 0.12 & -\\
Sr II  &$-$1.43    & $-$4.30  &  $-$0.80 & --       &   synth &(0.20)&  (0.12) & 0.23&-\\
Ba II &$-$1.72    & $-$3.90  &  $-$0.40 &  --       &  synth &(0.20)& (0.12)  & 0.23 & -\\
 \hline
\end{tabular}
\label{tab:ab}
\end{minipage}
\end{table*}

 For some abundances  in Table ~\ref{tab:ab}
 the standard variation ($\sigma_{\epsilon} $, column 5) involving small 
numbers of lines (column 6) is implausibly small. Therefore, 
for the abundances measured with less than five lines  we assume
as random error (column 7)  the largest standard variation 
obtained for the remaining abundances  (0.11 dex for Fe~II). 
The total error is obtained by quadratically
adding this updated random error with the systematic error.
To calculate the systematic errors, we have re-computed
the abundances, varying the model atmosphere considering
these uncertainties: $\Delta$ T$_{eff}$$\pm$ 100K,  $\Delta$ log(g)
$\pm$ 0.3 dex, $\Delta$ [Fe/H] $\pm$0.3 dex
and $\Delta$ \vmic $\pm$ 0.3 km s$^{-1}$.
 This  method to calculate the systematic errors 
and the uncertainties adopted are based on 
the recent work on EMP halo stars by  \citet{Yong13}; 
however similar method and values are used also 
in the work by \citet{Roederer14}.
The quadratic sum of these variations is reported in Table~\ref{tab:ab}.

\begin{figure*}[ht!]
\begin{minipage}{185mm}

\includegraphics[width=185mm]{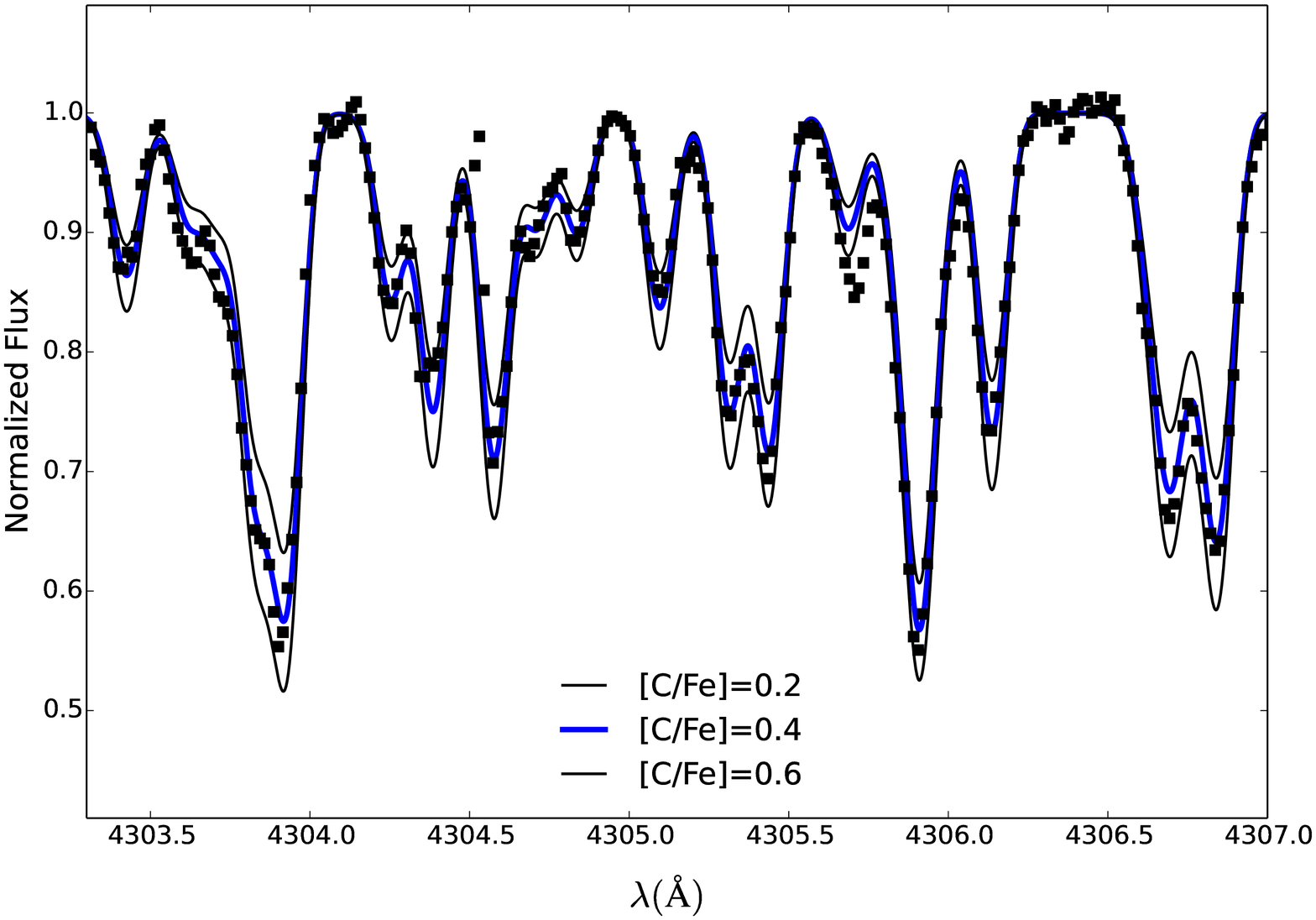}

\caption{Fit of CH lines of the G band in TYC 8442$-$1036$-$1. Dots:
observations, lines: synthetic spectra computed for the abundances
indicated}\label{fig1a}
\end{minipage}
\end{figure*}

We underline that the analysis performed in this work was made under
the assumption of 1D LTE.  Assuming LTE, means to imply that the
energy distribution is performed only by particles collision. This
assumption stops to be true close to the stellar surface (Bergemann \&
Nordlander 2014).  TYC 8442$-$1036$-$1 is a giant and extremely metal
poor star and the non-LTE effects can be important in particular for 
Na, Mg and Fe. In Table~\ref{tab:ab}, we report
  non-LTE corrections for these elements taken from literature, to give
  the reader an indication of the expected non-LTE effects.  We 
  find a variation up to +0.10 for Mg on two of the three lines
  considered here \citep[see][]{Osorio16}. However, following
  \citet{Merle11}, the corrections for Mg abundance are +0.19 dex for
  the line at 4703$\AA$ and 0.25 dex for the one at 5528 $\AA$. An
  intermediate results is obtained with the corrections by
  \citet{Mashonkina13}.  We also note that we do not take into
  account the line at 4571$\AA$ for which the correction could have
  been higher (up to 0.3 dex). For an estime of the non-LTE correction
  of Na, we provide the corrections calculated by the online 
  database INSPECT \footnote{http://inspect.coolstars19.com, A
    database for Interactive NLTE Spectroscopy of late-type stars.}; 
the corrections for Na in this database  are from \citet{Lind11}.
 Finally, we expect a non-LTE
  corrections for our Fe I of 0.1 dex (0.2 dex at maximum), based on
  \citet{Lind12}.  The abundance of Fe II is expected to be not
  affected by the non-LTE corrections and its value is still
  compatible with the estimate value of Fe I with non-LTE
  corrections. The corrections for non-LTE effect can be important
for Mn and Cr \citep{BG08,Bergemann10}; however, concerning Mn, recent
results obtained by \citet{Sneden15} can challenge the impact of
non-LTE corrections.  In Sect. 6 we present figures with the abundances
of TYC 8442$-$1036$-$1 and other observational data taken from literature.
In these figures, we use the abundances without non-LTE corrections, since
considering these corrections will not alter our conclusions
and most of the other data do not consider these corrections.

Abundances of C, Sr and Ba are derived using the spectral 
synthesis module of MOOG. The abundances of
 the elements are iteratively varied until the synthetic
 spectrum matched the observed one by visual inspection. The
macro-turbulent broadening is determined using a Gaussian
representing the combined effects of the instrumental profile,
atmospheric turbulence, and stellar rotation. The width of this
Gaussian is estimated during  the spectrum
synthesis fitting, and the abundances are thus (slightly)
sensitive to the adopted broadening.
The lines of Ba II are affected by hyperfine splitting and also 
by isotopic splitting. Therefore, the Ba abundances are computed
assuming the \citet{MCW98} r-process isotopic composition and
hyperfine splitting. We compare in Fig.~\ref{fig1c} the differences 
arising in the shape of the synthetic line of Ba at 4554\,$\AA$ when 
two different isotopic compositions are adopted:  
a solar compositions and a typical s-process composition. 
We note that the final abundance obtained from the lines 4554\,$\AA$ and 4914\,$\AA$ are 
about 0.4 dex lower taking the hyperfine splitting and isotopic splitting into
account. On the contrary, the abundance calculated on the lines 6110\,$\AA$
and 5883\,$\AA$ are in good agreement with the abundances measured from their EWs.
This was expected, given the lower impact of the splitting in the latter lines.
Also for the measurement of  Sr abundance we decide to use the synthetic analysis
(see Fig.~\ref{fig1b}).
The C abundance was measured in a similar way from analysis 
of the band of the A$-$X electronic
transitions of the CH molecule. Fig. \ref{fig1a} shows a comparison
of the synthetic spectrum with the data in the range 4303--4307\,$\AA$. 
The random abundance errors obtained with synthetic spectra (C, Sr and Ba)
are assumed to be 0.2 dex; in fact, for the three analysed elements
 the observed spectrum is inside this variation of the synthetic spectrum
(see Fig.~\ref{fig1b} and \ref{fig1a})
 We also assume for the systematic errors of these abundance 0.12 dex,
the maximum systematic error computed among the other elements, 
for a total error of 0.23 dex.

\section {Chemical evolution model}
The chemical evolution model used is the same as in \citet{Cescutti14}, 
which is based on the stochastic model developed in
\citet{Cesc08}, but with a different treatment of the gas flows, following
 the homogeneous model of \cite{Chiappini08}. 
The halo is assumed to consist of many independent regions, each with
the same typical volume, and each region does not interact with the
others.  Accordingly, the dimensions of the volume are  expected to be large 
enough to allow us to neglect the interactions between different volumes, 
at least as a first approximation. For typical ISM densities, a
supernova remnant becomes indistinguishable from the ISM -- that is, it
merges with the ISM -- before reaching $\sim50$ pc \citep{Thornton98};
 therefore, we decided to have a typical volume with a radius of roughly
90 pc.  The dimension of this volume is the same as in our previous works
adopting a stochastic model for the Galactic halo 
\citep{Cesc08,Cesc10,Cescutti13,Cescutti14,Cescutti15}.
 The number of assumed volumes to ensure a good
statistics in our previous models was 100; however given the
variation we implement here for the iron yields (see next section),
the new models are based on the results of 1000 volumes. 
 We did not use larger volumes because
they would produce more homogeneous results; in fact, in larger volumes
 the model would predict more SNe events and  a mixture 
of enrichments that would decrease the maximum spread
possible for the set of yields used.
Knowing the mass that is transformed into stars in a time step
(hereafter, $\mathrm{M_{stars}^{new}}$), we assigned the mass to one
star with a random function, weighted according to the initial mass
function (IMF) of \citet{Scalo86} in the range between 0.1 and
100~$M_{\odot}$.  We then extracted the mass of another star and
repeated this cycle until the total mass of newly formed stars
exceeded $\mathrm{M_{stars}^{new}}$.  In this way, $\mathrm{M_{stars}^{new}}$ is the
same in each region at each time step, but the total number and mass
distribution of the stars are different. We thus know the mass of each
star contained in each region, when it is born and when it will die,
assuming the stellar lifetimes of \citet{MM89}.  At the end of its
lifetime, each star enriches the ISM with its newly produced chemical
elements and with the elements locked in that star when it was formed,
excluding the fractions of the elements that are permanently locked
in to the remnant.
As shown in \citet{Cescutti13}, our model is able to reproduce the MDF
measured for the halo by \citet{Li10}. This comparison shows that the timescale of
enrichment of the model is compatible with  that of the halo stars
in the Solar vicinity. Moreover, our model predicts a small spread for
the $\alpha$-elements Ca and Si, 
which is compatible with the observational data.

\section {Modelling the nucleosynthesis}
\subsection{Stellar yields for Fe}

Our goal is to explore the impact on the chemical evolution model
 of the scenario in which massive stars do not always explode as SNe II
with a standard energy of 10$^{51}$erg, but they also 
explode with fainter explosions,  as observationally motivated by \citet{Moriya10}.

At the present, the mechanism of explosion of SNe II in not fully
understood \citep[see][]{Janka12} as well as possible connection
between mass and explosion energy for a SNe. Therefore, in the
nucleosynthesis results, the explosions are not obtained from first
principles, and they must be tuned in some way, typically given final
kinetic energy of the ejecta or a given amount of Fe ejected
\citep[see][]{Chieffi13,WW95,Nomoto06}.

In our model rather than stochastically select an
explosion energy and calculate the Fe ejected, we vary directly the Fe ejected.
In particular,  we decide to assume for the production of iron in  massive
stars (8 -- 80$M_{\odot}$) a distribution of yields which goes from
almost zero - 10$^{-5} M_{\odot}$ in the case of the faintest
explosions - to 0.2 $M_{\odot}$.  In this range, any value has the same probability to be
randomly chosen, so on average a massive star enriches the ISM with 0.1
$M_{\odot}$ of iron; in this way, the mean chemical evolution of
Fe is preserved. 

These assumptions are crude, but given the complexity connected to the
process of the explosion of a SNe II, we decide to keep our
assumptions as simple as possible. With this hypothesis, we can check
in our stochastic model the impact of the presence of a distribution
of energies from faint SNe to normal SNe.  In Sect. \ref{Results}, we
show for comparison the results obtained in \citet{Cescutti14}, with
our standard assumptions for Fe: the solar metallicity yields of
\citet{WW95}.  In all models, we have considered the SNe Ia
enrichment, as in \citet{Cesc06}.

\subsection{Stellar yields for C} 

For carbon, we present the results with two set of yields for the
massive stars and low-intermediate mass stars.  We have chosen these sets
to visualize the difference between the carbon production in rotating
and non-rotating stellar evolution models:

\begin{itemize}
\item \emph{rotating yields}, the set of yields of the model a described in
  \citet{Cesc10}, based on the yields by \citet{Meynet02} for z~$\ge
  10^{-5}$, and on the total yields by \citet{Hirschi07} for z~=~$10^{-8}$.
For low-intermediate mass stars, we adopt the stellar yields by
\citet{Meynet02}.
\item \emph{non-rotating yields}, the carbon yields calculated by
  \citet{WW95}, which is a set of yields frequently adopted in chemical
  evolution studies, but without rotation. For the low-intermediate
  mass stars we assume the yields by \citet{vandenHoek97}.
\end{itemize}

The two set of yields for carbon do not originate from the same
group and/or stellar evolution code, so it is possible that also
systematic effects produce differences between them 
and not only the rotation. Nevertheless, we confirm that
yields from the Geneve group without rotation produce less 
carbon compared to  the \emph{rotating yields}, similarly to 
\emph{non-rotating yields} assumed here.

\subsection{Stellar yields for Ba and Sr} 
For barium and strontium, we use for all the models with rotating
massive stars the nucleosynthesis of the \emph{MRD+s B2} model
described in \citet{Cescutti14}.  These elements can be produced in
this model by both the s-process and the r-process in massive
stars. The assumed r-process scenario follows the idea described in
\citet{Winteler12} and recently confirmed by \citet{Nishimura15},
where a small percentage of massive stars end their lives as
magneto-rotationally driven (MRD) SNe.   To implement this
  scenario into our chemical evolution model, we randomly select 10\%
  of all the simulated massive stars and we assume that these massive
  stars generate an r-process event at the end of their lives.  We
  have no prediction of the ejected mass in each r-process event.  On
  these grounds, we assume that the MRD scenario produces the same
  amount of Ba in a stellar generation as the EC+s model
  \citep{Cescutti13}; these empirical yields were obtained as the
  simplest array able to reproduce the observed trend in Galactic halo
  star of increasing [Ba/Fe] with metallicity. In this scenario, we
take also into account the possibility that the amount of r-process
material ejected is not constant \citep[for details on the variation
see][]{Cescutti14}.  The presence or absence of rotation does not
influence the r-process production in our set of yields. The
contribution by the s-process in rotating massive stars is assumed as
in the \emph{fs-}model of \citet{Cescutti13}, where we considered the
stellar yields obtained by \citet{Frisch12}.  The barium and strontium
produced by the s-process is only barely affected by the SNe II
explosion, and therefore it is relatively safe to consider a variation
of iron without changing these yields. It may be not the case for the
Ba r-process production, and we comment on this in the Section
\ref{Results}.  In the non-rotating model there is no s-process
production of Ba and Sr from rotating massive stars.  However, in all
the models, we consider the s-process contribution from stars in the
mass range 1.3--3~$M_{\odot}$, by implementing the yields by
\citet{Cristallo09,Cristallo11}. We underline that this production
channel affects the model results only at moderate metallicity
([Fe/H]~$\sim -$1.5 dex).

\section {Results}\label{Results}
\subsection{Results for Ba} 
\begin{table}[!]
\caption{The nucleosynthesis prescriptions adopted for the 3 models.}\label{tabY}
\begin{tabular}{|c|c|c|}
\hline
Models Fig.~\ref{fig2}, \ref{fig2b} \& \ref{fig3}    & yields for Fe & yields for Ba and C \\
\hline
\hline
spinstars scenario                  & standard & rotating    \\
\hline
 faint SNe scenario           & with variance & non-rotating \\
\hline
 spinning faint scenario & with variance & rotating    \\
\hline
\end{tabular}
\end{table}

\begin{figure*}[ht!]
\begin{minipage}{185mm}
\vspace{-4cm}
\includegraphics[width=185mm]{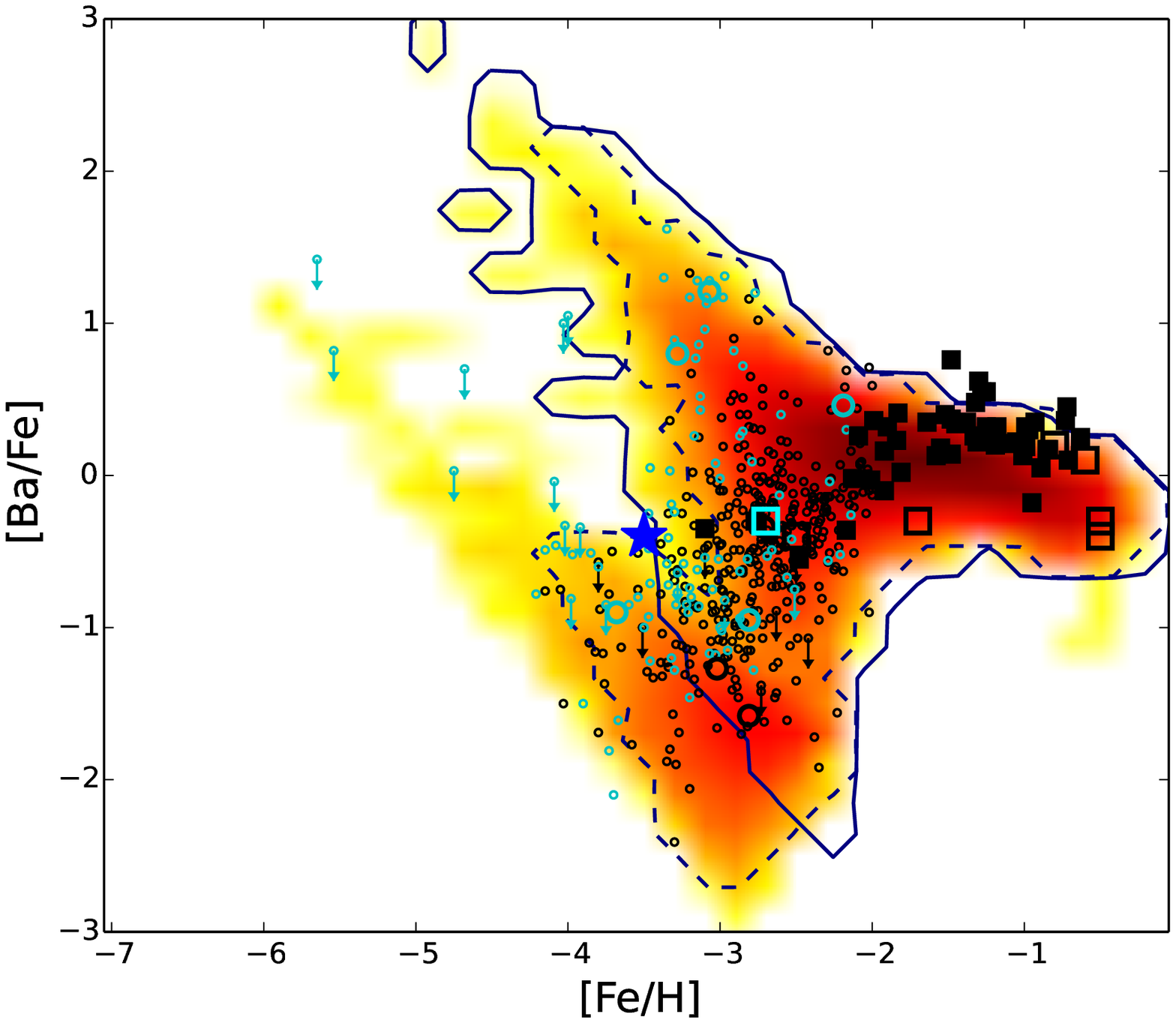}
\vspace{-2cm}
\begin{center}
\includegraphics[width=160mm]{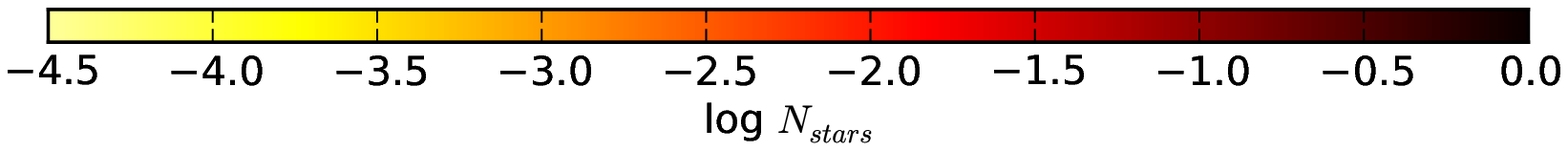}
\end{center}

\caption{[Ba/Fe] vs. [Fe/H]. The star analysed here is shown using a
  blue star; the open small dots are data collected by
  \citet{Placco14} and the arrows indicate upper limits; the big open
  dots are data from \citet{HansenT15}; the solid squares from
  \citet{Hansen12}; the open squares are taken from \citet{Hansen15}.
  The cyan symbols refer to CEMP-no stars, adopting the criterion
  [C/Fe]~$>$~0.7 condition for carbon enhancement, as in
  \citet{Placco14}; the black symbols are for normal star. We do not
  plot stars with s-process enhancement (CEMP-s).  The color-coded
  surface density plot presents the density of long living stars for
  the model of spinning faint SNe.  The contour plot with dashed line
  shows the results for the model of spinstars, the contour plot with
  solid line shows the results for the model of faint
  SNe. }\label{fig2}
\end{minipage}
\end{figure*}

 In Fig.~\ref{fig2}, we show the results of three  models for [Ba/Fe] vs.
[Fe/H]. 
Two models take into account  rotating massive stars:
one model considers the presence of faint SNe that
can produce an almost negligible amount of iron (``spinning faint scenario''),  
in the other (contour plot with dashed line), the SNe II produce a fixed amount of iron (roughly
0.1 M$_{\odot}$), following the results by \citet{WW95} (``spinstar scenario''). 
In the last model (contour plot with solid line), we consider non-rotating massive stars and
the presence of faint SNe (``faint SNe scenario''). The
nucleosynthesis yields are summarised in Table~\ref{tabY}.  We compare
our models to the star analysed here and to a 
collection of observational data.

The ``spinstars scenario'' \citep[that is the same model for Ba as the
model ``MDR+s B2'' realised in][]{Cescutti14} is quite successful,
because in the space [Ba/Fe] vs. [Fe/H], the density of its simulated
long-living stars is matching most of the stars observed in the halo.
However, a certain number of objects at extremely low
metallicity and low [Ba/Fe] are positioned where the model  predicts
a null density of stars. It was also unable to explain stars at
[Fe/H]~$\sim -$3.5 dex at [Ba/Fe]~$\sim -$0.5 dex.  In particular also
the star we have characterised here is found in this area.

\begin{figure*}[ht!]
\begin{minipage}{185mm}
\vspace{-4cm}
\includegraphics[width=185mm]{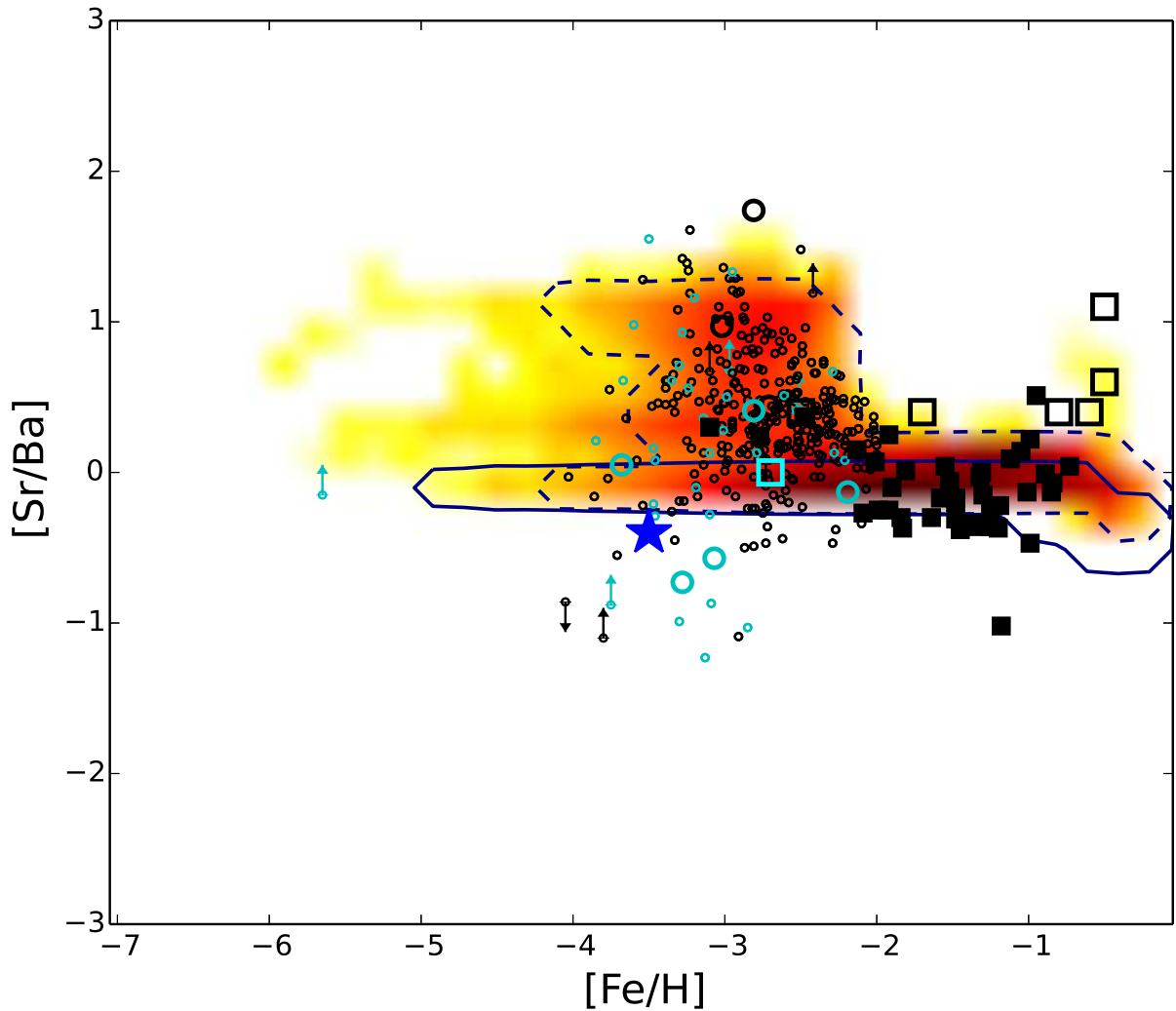}
\vspace{-2cm}
\caption{[Sr/Ba] vs. [Fe/H]. The symbols are the same as in Fig.~\ref{fig2}.
The color-coded surface density plot and the contour plots describe the same models 
as in Fig.~\ref{fig2}.}\label{fig2b}
\end{minipage}
\end{figure*}

On the other hand, in Fig.~\ref{fig2}, the results of the model ``spinning faint
scenario'' explain, within the observational errors all the stars at
[Fe/H]~$< -$4 dex.  It also simulates stars at [Fe/H]~$\sim -$3.5
dex at [Ba/Fe]~$\sim -$0.5 dex and compatible with the star analysed
here. 
The density plot produced by this model shows two bands which
move downwards from [Fe/H]~$\sim-$6 dex to [Fe/H]~$\sim -$3 dex. The
band at lower [Ba/Fe] has been enriched by spinstars with an
associated SNe II with a low iron production (faint SNe).  Therefore,
this coupling helps in better recovering the observed stars at the
lowest [Fe/H], which are mostly in this lower band.  The second band
with low [Fe/H] but high [Ba/Fe] is produced in the model by r-process
events coupled with weak production of iron.  Stars in the region are
not observed for [Fe/H]~$< -$4 dex; if this absence will be confirmed
by future observations, it will provide an additional contraint to the
r-process events: they should be associated with a normal production
of Fe, and not to faint SNe.  This constraint applies in case of
single massive stars as progenitors of the r-process events, whereas
in the case of neutron-stars merger scenario, it should be applied to
at least one of the two massive progenitor stars.  

In the model  ``faint SNe scenario'', we do not consider production of s-process
  by massive stars.  The results of this model are only marginally consistent
  with the abundances measured in TYC 8442$-$1036$-$1. Considering
non-LTE corrections for our star will increase its agreement with the
``faint SNe scenario''; however, this model still fails to reproduce the
 stars  located in the band at lower [Ba/Fe] in the results of
  the ``spinning faint scenario''.

\subsection{Results for [Sr/Ba]} 
We present in the Fig.\ref{fig2b}, the results of the three  models for [Sr/Ba] vs.
[Fe/H]. Similarly to the [Ba/Fe] case in the previous section, the``spinstars
scenario'' model is really successful, because in the space [Sr/Ba]
vs.  [Fe/H], the density of its simulated
long-living stars can recover most of the stars observed in the halo.
Still, a certain number of objects are located at extremely low metallicity 
where this model predicts a null density of stars.
Again this issue is improved, once we adopted also a variation of
yield for the iron, as in the case of of ``spinning faint
scenario'' .

The Fig. \ref{fig2b} also explained why the spinstars s-process
contribution is essential. Indeed the ``faint SNe scenario'' - without
this contribution - cannot reproduce a large fraction of abundances
observed in Galactic stars.  We note that in this figure there is also
fraction of stars that is compatible with none of the scenarios
investigated here.  These outliers are located below the [Sr/Ba] ratio
assumed for the r-process events and cannot be reproduced, being also
the s-process produced by the spinstars with a [Sr/Ba]$>$0.  However,
their fraction is small and in the [Sr/Ba] plot the errors of both
chemical element abundances should be considered; therefore a
substantial fraction of them is still compatible within the errors to
the results and TYC 8442$-$1036$-$1 is in this group.  Finally,the
CEMP-no stars in this group could be originated in a binary system,
therefore they show a mild enhancement due to the pollution of
s-process material from the companion star.  If this possible
scenario is taken into account only two EMP stars are outliers
compared to our models which is an excellent results.

\begin{figure*}[ht!]
\begin{minipage}{185mm}
\vspace{-4cm}
\includegraphics[width=185mm]{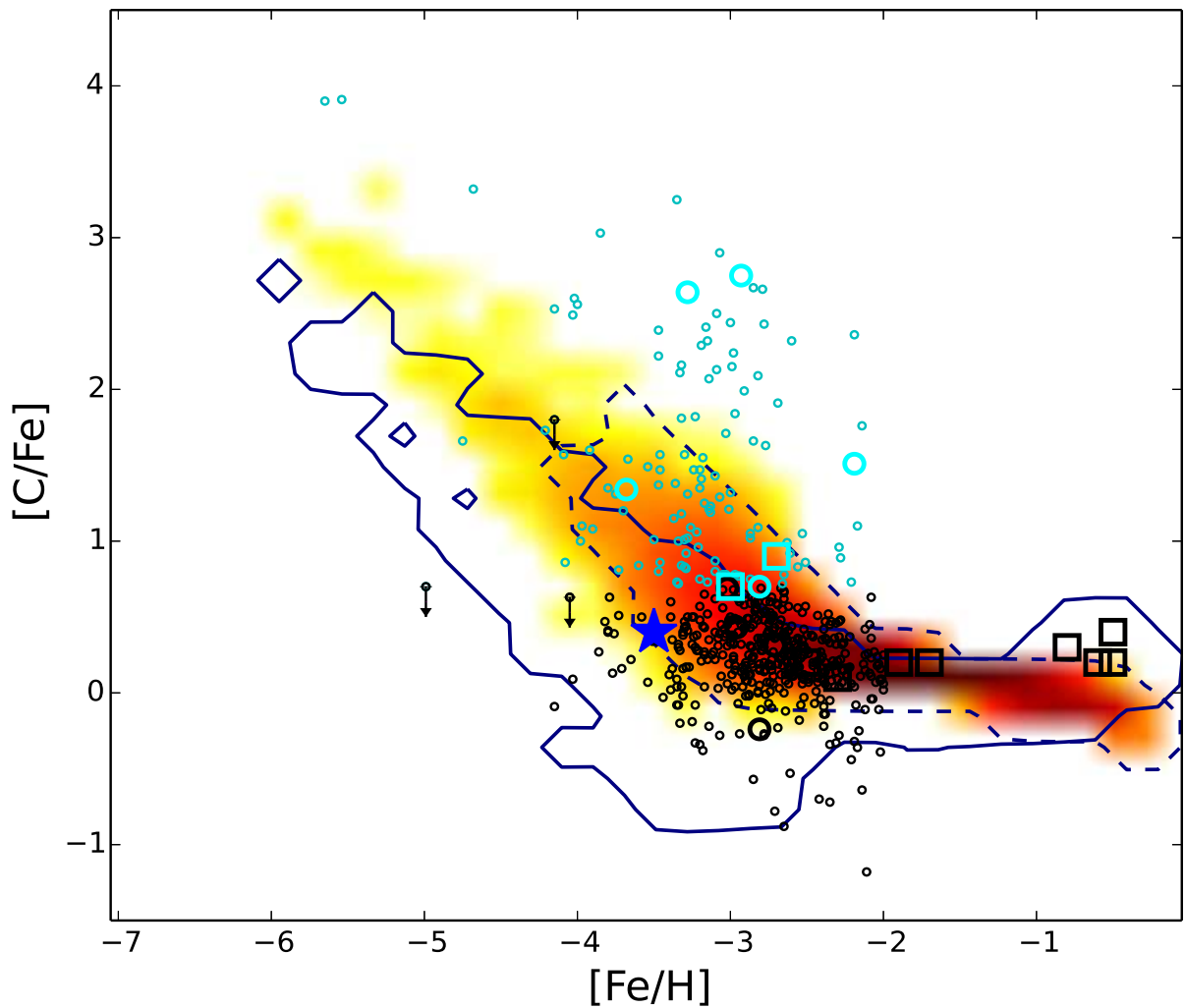}
\vspace{-2cm}
\caption{[C/Fe] vs. [Fe/H]. The symbols are the same as in Fig.~\ref{fig2}.
The color-coded surface density plot and the contour plots describe the same models 
as in Fig.~\ref{fig2}.}\label{fig3}
\end{minipage}
\end{figure*}

\subsection{Results for C} 

Going towards extremely low metallicity, an increasing fraction of stars
are carbon-rich and belong to the category of the CEMP-no stars \footnote{
In \citet{Beers05}, a metal-poor star is a CEMP star if [C/Fe]$>$1.
If there is no excess of s-process ([Ba/Fe]$<1$) the star belongs to 
the CEMP-no category, otherwise to the CEMP-s class.}.
However, the star we have analysed does
not belong to this category, being only slightly carbon enhanced.
Therefore, we decide to investigate how our  spinning faint SNe
model behaves in terms of [C/Fe] ratio vs. [Fe/H]. The results are
shown in Fig.~\ref{fig3}.   In this figure again the density plot
represents the results for the ``spinning faint SNe scenario'', whereas the
contour plot with dashed line shows the results of  the ``spinstars scenario'' 
where SNe II produce a fixed amount of iron. The solid line contour plot 
represents the results assuming non-rotating yields for
carbon (see Table~\ref{tabY} for details), and in this 
model (``faint SNe scenario'') we consider a variation in iron yields.

The low production of iron by faint SNe produces a rise in the [C/Fe]
ratio towards low metallicity, and this trend is common in both models
that consider faint SNe (``spinning faint SNe scenario'' and ``faint SNe scenario'')
with and without rotation.  However, the model with rotation predicts
a density distribution of stars with a [C/Fe] ratio about 1 dex higher
compared to the non-rotating yields for [Fe/H]~$<-$3 dex.

The higher production of carbon in the yields with rotation improve
the agreement between the model and the data. In fact, a substantial
number of Galactic stars is inside the predictions of the faint
spinstars model for $-$3$\,<\,$[Fe/H]$<-$4 dex and [C/Fe]$\sim$1 dex,
and outside the predictions of the model assuming yields without
rotation.  The chemical enrichment of the star measured here, is
consistent with both rotation and non-rotational models.  We also note
that the model with rotation cannot explain a fraction of stars with
lower ratio of [C/Fe] that are possible to be reproduced in the model
without rotation. This could show a fact that is expected: a
distribution of rotational velocities among the massive stars.  We
underline that it will be also possible in the near future to
investigate the most likely distribution of stellar rotation for these
low-metallicity massive stars with a set of nucleosynthesis
computations where yields for different rotational velocity are
provided.   In comparison with the ``spinning faint SNe scenario''
  the model with fixed amount of iron produced by SNe II (``spinstars
  scenario'', dashed contour) cannot reproduce the data for
  [Fe/H]$<-$4~dex.  Moreover, a not negligible fraction of stars is
  located just inside the upper edge of the contour for
  $-$4$<$[Fe/H]$<-$3, where the results of the model predict a very
  low density of long living stars; therefore the model is not fully
  consistent with the data.

The spinning faint SNe model cannot be considered an exhaustive
explanation for the CEMP-no stars.  A not negligible fraction of the
observed CEMP-no stars present a [C/Fe] not compatible with the
predictions of the model.  The carbon present in these objects is in
some case more than 1 dex compared to the spinning faint SNe model
results.  This class of objects has also been identified in
\citet{Cooke14}, and called ``super CEMP''. 

It is likely that the carbon present (at least) in this class of
CEMP-no stars was not well mixed in the ISM, before being locked in
these low-mass stars. In \citet{Meynet10}, the chemical signatures
present in the three most iron-poor stars known (at that time),
\footnote{HE 0107$-$5240 \citep{Christlieb02}, HE 1327$-$2326
  \citep{Frebel05}, HE 0557$-$4840 \citep{Norris07}; the record is now
  held by SMSS J031300.36$-$670839.3 \citep{Keller14}. } which are also
CEMP-no stars, were explained assuming that these stars were formed
(almost) entirely by stellar winds of rotating massive stars.  In this
scenario the [C/Fe] of these stars can be strongly enhanced compared
to the results of a standard chemical model, where the stellar
enrichment is well mixed with the ISM before forming new stars. In our
plots, these stars are present, they belong to the category of stars
that lay above our model predictions and are the first three data
points starting approximately at [Fe/H]=~$-$6 dex.  The presence
  of this class of ``super CEMP'', was already noted in
  \citet{Cesc10}, where we could not reconcile our models with a large
  fraction of the CEMP-no (known at that time). The issue was
  quite clear also in the [C/O] (and [N/O]) vs [O/H] space, where it was
  possible to neglect the influence of the uncertainties in the
  production of iron. 

Recently, it was also underlined by \citet{Maeder15} that different
sub-classes of CEMP-no stars should be considered in the context of
formation by stellar winds. They are probably determined by different
degrees of internal mixing during stellar evolution.  It will be
possible to take all these differences into account only when models
with nucleosynthesis covering a broad range of stellar masses,
initial metallicities, and CNO ratios, rotational velocities, and mass
loss rates will be available. Therefore, it is not surprising that in
the context of a chemical evolution model, where only full mixing with
the ISM is considered and just a grid of models for two velocities are
available, we cannot fully explain all the CEMP-no stars.

Another possible explanation is that at least a fraction of these
CEMP-no stars are the secondary in  a binary system. In this scenario,
the star presents the pristine composition polluted by AGB material of
the primary star.  This is the same scenario that is favorite for the
CEMP-s, and in this case the absence of strong enrichment of barium
can be explained in the framework of classical theoretical yields for
AGB stars; the very low-metallicity reduces the barium production 
and enhances the production of the lead \citep{Cristallo11}. It is also
possible that the evolution of low-mass star is quite different
at extremely-low metallicity and it can suppress the s-process production
\citep{Fujimoto00, Komiya07}.  As found in \citet{Cooke14},
at least 3 out of 5 the stars of this group show evidence to be binary
\citep{Starkenburg14}; however, very recently \citet{HansenT15} found no
compelling relation between binarity and carbon enhancement.

Given the complexity of the observational data, we have shown
that the spinning faint SNe model has successfully recovered the main
trend of the data. In fact, the star studied here and a substantial fraction of
extremely metal-poor (CEMP-no and normal) stars can be formed in the
framework of normal chemical evolution, if we couple the fast-rotating yields
and the presence of faint SNe.  Moreover, for elements that
are not expected to be ejected by stellar winds, as Ba, we have shown
that basically all the observational data available are compatible
with the predictions of the spinning faint SNe model.

\section {Conclusions}

We have measured from a high-resolution spectrum 13 chemical elements
for TYC 8442$-$1036$-$1, a metal-poor star of the Galactic halo.  This
star belongs to the rare class of EMP stars with
[Fe/H]~=~$-$3.5$\pm0.13$ dex, 0.5 dex lower than what previously
determined with a medium resolution spectrum.  At this metallicity,
most of the stars in the Galactic halo have a [C/Fe]~$>$~0.7 dex and
belong to the class of the CEMP-no stars.  This is not the case for
our star which shows just mild overabundance of carbon ([C/Fe]~=~0.4
dex ).  We have also measured with particular attention its [Ba/Fe]
ratio, and we find a low abundance of about [Ba/Fe]~=~ $-$0.4 dex.
This particular abundance pattern was not explained by our previous
models for neutron capture elements in the Galactic halo. In our
previous work \citep{Cescutti13,Cescutti14}, we showed that an
r-process component and the spinstars contribution of s-process can
account for most of the data in literature.  We decide thus to include
also a variation on the iron yields mimicking the production of iron
by faint SNe. The final model, that we call spinning faint SNe, is
able to explain the presence also of the most extreme stars in the
[Ba/Fe] ratio vs. [Fe/H] space.  The comparison of the model with
  the observational data also indicates that the r-process events are
  not linked to faint SNe events.  Since most of the stars at such a
low metallicity appear to be quite enhanced in carbon, we decided to
show also the results of our model in the [C/Fe] ratio vs. [Fe/H]
space.  We find that the model is able to explain the chemistry of TYC
8442$-$1036$-$1 and also to recover a large fraction of the CEMP-no
stars.  However, a not negligible fraction still remains not
explicable. A scenario to explain these super CEMP-no stars is that
these stars have been formed (almost) entirely from the material
ejected through winds by fast rotating massive stars
\citep{Meynet10,Maeder15}.

\begin{acknowledgements} 
  This work was partially supported by the UK Science and Technology
  Facilities Council (grant ST/M000958/1). We thank the referee for
  comments that improved the clarity of this paper. GC thanks Lorenzo
  Monaco and Giacomo Beccari for the support during the execution of
  the observations. GC also thanks Raphael Hirschi for his useful
  suggestions, Marco Limongi for his comments on the iron production,
  Sean Ryan and Piercarlo Bonifacio for their advices on the section
  of data analysis.

\end{acknowledgements}

\bibliographystyle{aa}
\bibliography{spectro}
\newpage

\appendix
\begin{table}

\caption{Equivalent width for TYC 8442$-$1036$-$1. }

\begin{tabular}{|r|c|r|r|r|}

\hline
\hline
 Wavelength ($\AA$)&   Species  &   L.E.P. (eV)    & log $gf$   &     $W_{\lambda}$ (m$\AA$)      \\
\hline
  5889.951 &  Na I &   0.000 &  0.112  &  89.11 \\
  5895.924 &  Na I &   0.000 &$-$0.191  &  69.92 \\
  4351.906 &  Mg I &   4.340 &$-$0.525  &  29.42 \\
  4702.990 &  Mg I &   4.330 &$-$0.380  &  43.16 \\
  5528.405 &  Mg I &   4.340 &$-$0.341  &  42.12 \\
  4283.011 &  Ca I &   1.890 &$-$0.220  &  22.29 \\
  4318.652 &  Ca I &   1.900 &$-$0.210  &  18.21 \\
  4425.437 &  Ca I &   1.880 &$-$0.360  &  15.01 \\
  4435.679 &  Ca I &   1.890 &$-$0.520  &   8.57 \\
  4454.779 &  Ca I &   1.900 &  0.260  &  35.86 \\
  5265.556 &  Ca I &   2.520 &$-$0.260  &   5.54 \\
  5588.749 &  Ca I &   2.520 &  0.210  &  12.43 \\
  5857.451 &  Ca I &   2.930 &  0.230  &   4.99 \\
  6102.723 &  Ca I &   1.880 &$-$0.790  &   6.63 \\
  6122.217 &  Ca I &   1.890 &$-$0.320  &  17.35 \\
  6162.173 &  Ca I &   1.900 &$-$0.090  &  27.52 \\
  4246.822 &  Sc II &  0.310 &  0.240  &  83.71 \\
  4314.083 &  Sc II &  0.620 &$-$0.100  &  46.27 \\
  4400.389 &  Sc II &  0.610 &$-$0.540  &  29.74 \\
  4415.557 &  Sc II &  0.600 &$-$0.670  &  23.33 \\
  5031.021 &  Sc II &  1.360 &$-$0.400  &   8.18 \\
  5526.790 &  Sc II &  1.770 &  0.030  &   5.78 \\
  4533.241 &  Ti I &   0.850 &  0.480  &  16.02 \\
  4534.776 &  Ti I &   0.840 &  0.280  &  13.82 \\
  4981.731 &  Ti I &   0.840 &  0.500  &  20.44 \\
  4991.065 &  Ti I &   0.840 &  0.380  &  16.15 \\
  4999.503 &  Ti I &   0.830 &  0.250  &  11.85 \\
  5014.276 &  Ti I &   0.810 &  0.110  &   8.80 \\
  5039.957 &  Ti I &   0.020 &$-$1.130  &   4.20 \\
  5064.653 &  Ti I &   0.050 &$-$0.990  &   4.67 \\
  4290.219 &  Ti II &   1.160& $-$0.930 &   56.93 \\
  4300.049 &  Ti II &   1.180& $-$0.490 &   61.91 \\
  4337.915 &  Ti II &   1.080& $-$0.980 &   53.50 \\
  4394.051 &  Ti II &   1.220& $-$1.770 &   12.88 \\
  4395.033 &  Ti II &   1.080& $-$0.510 &   73.55 \\
  4395.850 &  Ti II &   1.240& $-$1.970 &    7.46 \\
  4399.772 &  Ti II &   1.240& $-$1.220 &   38.05 \\
  4417.719 &  Ti II &   1.160& $-$1.230 &   42.94 \\
  4418.330 &  Ti II &   1.240& $-$1.990 &    6.96 \\
  4443.794 &  Ti II &   1.080& $-$0.700 &   74.70 \\
  4444.558 &  Ti II &   1.120& $-$2.210 &   10.41 \\
  4450.482 &  Ti II &   1.080& $-$1.510 &   31.54 \\
  4464.450 &  Ti II &   1.160& $-$1.810 &   15.35 \\
  4468.507 &  Ti II &   1.130& $-$0.600 &   68.69 \\
  4470.857 &  Ti II &   1.160& $-$2.060 &    7.40 \\
  4501.273 &  Ti II &   1.120& $-$0.760 &   66.05 \\
  4533.969 &  Ti II &   1.240& $-$0.540 &   67.72 \\
  4563.761 &  Ti II &   1.220& $-$0.790 &   58.07 \\
  4571.968 &  Ti II &   1.570& $-$0.230 &   57.79 \\
  4657.200 &  Ti II &   1.240& $-$2.240 &    4.38 \\
  5129.152 &  Ti II &   1.890& $-$1.300 &    6.00 \\
  5336.771 &  Ti II &   1.580& $-$1.630 &    9.19 \\
  5381.015 &  Ti II &   1.570& $-$1.970 &    5.72 \\
  4379.230 &  V I &   0.300  & 0.550   &   5.11 \\
  4254.332 &  Cr I &   0.000 &$-$0.110  &  66.85 \\
  4274.796 &  Cr I &   0.000 &$-$0.230  &  60.79 \\
  4289.716 &  Cr I &   0.000 &$-$0.360  &  55.05 \\
  5409.772 &  Cr I &   1.030 &$-$0.720  &   6.65 \\
  4754.042 &  Mn I &  2.280  &-0.090   &   5.34 \\
  4823.524 &  Mn I &  2.320  & 0.140   &   6.50 \\
\hline
\hline
\end{tabular}
\label{tab:ew1}
\end{table}

\begin{table}

\vspace{0.8cm}
\begin{tabular}{|r|c|r|r|r|}

\hline
\hline
 Wavelength ($\AA$)&   Species  &   L.E.P. (eV)    & log $gf$   &     $W_{\lambda}$ (m$\AA$)      \\
\hline

  4184.890 &  Fe I &   2.830 &$-$0.870  &  14.21 \\
  4187.039 &  Fe I &   2.450 &$-$0.550  &  47.95 \\
  4187.795 &  Fe I &   2.420 &$-$0.550  &  51.02 \\
  4191.431 &  Fe I &   2.470 &$-$0.730  &  37.05 \\
  4195.329 &  Fe I &   3.330 &$-$0.410  &  11.11 \\
  4199.095 &  Fe I &   3.050 &  0.250  &  46.63 \\
  4202.029 &  Fe I &   1.480 &$-$0.700  &  86.50 \\
  4222.213 &  Fe I &   2.450 &$-$0.970  &  30.57 \\
  4227.427 &  Fe I &   3.330 &  0.230  &  40.13 \\
  4233.603 &  Fe I &   2.480 &$-$0.600  &  40.67 \\
  4238.810 &  Fe I &   3.400 &$-$0.230  &  19.09 \\
  4250.119 &  Fe I &   2.470 &$-$0.400  &  50.75 \\
  4260.474 &  Fe I &   2.400 &$-$0.020  &  76.35 \\
  4271.154 &  Fe I &   2.450 &$-$0.350  &  57.11 \\
  4282.403 &  Fe I &   2.170 &$-$0.820  &  44.59 \\
  4337.046 &  Fe I &   1.560 &$-$1.700  &  43.97 \\
  4415.123 &  Fe I &   1.610 &$-$0.610  &  85.44 \\
  4430.614 &  Fe I &   2.220 &$-$1.660  &  12.55 \\
  4442.339 &  Fe I &   2.200 &$-$1.250  &  27.52 \\
  4447.717 &  Fe I &   2.220 &$-$1.340  &  25.53 \\
  4461.653 &  Fe I &   0.090 &$-$3.200  &  60.88 \\
  4466.552 &  Fe I &   2.830 &$-$0.600  &  30.33 \\
  4489.739 &  Fe I &   0.120 &$-$3.970  &  20.88 \\
  4494.563 &  Fe I &   2.200 &$-$1.140  &  35.54 \\
  4528.614 &  Fe I &   2.180 &$-$0.820  &  51.25 \\
  4531.148 &  Fe I &   1.480 &$-$2.150  &  26.30 \\
  4592.650 &  Fe I &   1.560 &$-$2.460  &  13.65 \\
  4602.940 &  Fe I &   1.490 &$-$2.210  &  26.46 \\
  4647.430 &  Fe I &   2.950 &$-$1.350  &   4.34 \\
  4871.318 &  Fe I &   2.870 &$-$0.360  &  35.08 \\
  4872.138 &  Fe I &   2.880 &$-$0.570  &  24.98 \\
  4891.492 &  Fe I &   2.850 &$-$0.110  &  49.26 \\
  4903.310 &  Fe I &   2.880 &$-$0.930  &  14.44 \\
  4918.994 &  Fe I &   2.870 &$-$0.340  &  36.83 \\
  4920.503 &  Fe I &   2.830 &  0.070  &  55.96 \\
  4938.814 &  Fe I &   2.870 &$-$1.080  &   8.90 \\
  4939.687 &  Fe I &   0.860 &$-$3.340  &  14.13 \\
  4994.130 &  Fe I &   0.920 &$-$3.080  &  20.67 \\
  5001.864 &  Fe I &   3.880 &  0.010  &   7.40 \\
  5006.119 &  Fe I &   2.830 &$-$0.620  &  23.36 \\
  5041.072 &  Fe I &   0.960 &$-$3.090  &  17.71 \\
  5041.756 &  Fe I &   1.490 &$-$2.200  &  24.04 \\
  5049.820 &  Fe I &   2.280 &$-$1.360  &  23.78 \\
  5051.635 &  Fe I &   0.920 &$-$2.800  &  32.09 \\
  5068.766 &  Fe I &   2.940 &$-$1.040  &   7.63 \\
  5079.740 &  Fe I &   0.990 &$-$3.220  &  12.20 \\
  5083.339 &  Fe I &   0.960 &$-$2.960  &  22.72 \\
  5110.413 &  Fe I &   0.000 &$-$3.760  &  41.65 \\
  5123.720 &  Fe I &   1.010 &$-$3.070  &  17.15 \\
  5127.359 &  Fe I &   0.920 &$-$3.310  &  10.90 \\
  5131.470 &  Fe I &   2.220 &$-$2.520  &   2.60 \\
  5254.955 &  Fe I &   0.110 &$-$4.760  &   4.74 \\
  5266.555 &  Fe I &   3.000 &$-$0.390  &  27.07 \\
  5269.537 &  Fe I &   0.860 &$-$1.320  & 106.40 \\
  5281.790 &  Fe I &   3.040 &$-$0.830  &  11.96 \\
  5283.621 &  Fe I &   3.240 &$-$0.520  &  13.52 \\
  5302.302 &  Fe I &   3.280 &$-$0.880  &   7.25 \\
  5307.361 &  Fe I &   1.610 &$-$2.990  &   4.02 \\
  5324.179 &  Fe I &   3.210 &$-$0.240  &  28.92 \\
  5328.039 &  Fe I &   0.920 &$-$1.470  &  97.30 \\
\hline
\hline
\end{tabular}
\label{tab:ew2}
\end{table}

 \begin{table}

\begin{tabular}{|r|c|r|r|r|}

\hline
\hline
 Wavelength ($\AA$)&   Species  &   L.E.P. (eV)    & log $gf$   &     $W_{\lambda}$ (m$\AA$)      \\
\hline

  5328.532 &  Fe I &   1.560 &$-$1.850  &  42.70 \\
  5339.929 &  Fe I &   3.270 &$-$0.720  &   9.95 \\
  5367.470 &  Fe I &   4.420 &  0.440  &   4.88 \\
  5369.962 &  Fe I &   4.370 &  0.540  &   8.29 \\
  5371.490 &  Fe I &   0.960 &$-$1.650  &  91.01 \\
  5383.369 &  Fe I &   4.310 &  0.640  &  15.37 \\
  5393.168 &  Fe I &   3.240 &$-$0.910  &   8.29 \\
  5397.128 &  Fe I &   0.920 &$-$1.990  &  78.19 \\
  5405.775 &  Fe I &   0.990 &$-$1.840  &  80.86 \\
  5410.910 &  Fe I &   4.470 &  0.400  &   4.34 \\
  5415.200 &  Fe I &   4.390 &  0.640  &  11.78 \\
  5429.697 &  Fe I &   0.960 &$-$1.880  &  82.44 \\
  5434.524 &  Fe I &   1.010 &$-$2.120  &  65.98 \\
  5446.917 &  Fe I &   0.990 &$-$1.910  &  77.60 \\
  5455.609 &  Fe I &   1.010 &$-$2.090  &  70.81 \\
  5497.516 &  Fe I &   1.010 &$-$2.850  &  26.69 \\
  5501.465 &  Fe I &   0.960 &$-$3.050  &  22.00 \\
  5506.779 &  Fe I &   0.990 &$-$2.800  &  29.13 \\
  5569.618 &  Fe I &   3.420 &$-$0.540  &  10.23 \\
  5572.842 &  Fe I &   3.400 &$-$0.310  &  15.23 \\
  5576.089 &  Fe I &   3.430 &$-$1.000  &   3.27 \\
  5586.756 &  Fe I &   3.370 &$-$0.140  &  21.36 \\
  5615.644 &  Fe I &   3.330 &$-$0.140  &  30.07 \\
  6065.480 &  Fe I &   2.610 &$-$1.410  &   9.84 \\
  6136.615 &  Fe I &   2.450 &$-$1.400  &  17.43 \\
  6137.692 &  Fe I &   2.590 &$-$1.400  &  12.11 \\
  4233.172 &  Fe II &  2.580 &$-$1.900  &  36.10 \\
  4416.830 &  Fe II &  2.780 &$-$2.410  &   7.66 \\
  4491.405 &  Fe II &  2.860 &$-$2.700  &   7.20 \\
  4508.288 &  Fe II &  2.860 &$-$2.250  &  13.71 \\
  4515.339 &  Fe II &  2.840 &$-$2.450  &   9.79 \\
  4520.224 &  Fe II &  2.810 &$-$2.600  &   8.70 \\
  4522.634 &  Fe II &  2.840 &$-$2.030  &  21.74 \\
  4541.524 &  Fe II &  2.860 &$-$2.790  &   4.39 \\
  4555.893 &  Fe II &  2.830 &$-$2.160  &  13.09 \\
  4576.340 &  Fe II &  2.840 &$-$2.820  &   3.84 \\
  5234.625 &  Fe II &  3.220 &$-$2.150  &  11.24 \\
  5476.900 &  Ni I  &  1.830   & $-$0.890 &   40.62\\
  4722.153 &  Zn I &   4.030 &$-$0.338  &    3.42 \\ 
  4810.528 &  Zn I &   4.080 &$-$0.137  &    3.57 \\
\hline
\hline
\end{tabular}
\label{tab:ew3}
\end{table}

\end{document}